# When Peers Outperform AI (and When They Don't): Interaction Quality Over Modality

Caitlin Morris, MIT Media Lab, camorris@media.mit.edu
Pattie Maes, MIT Media Lab

**Abstract:** As AI increasingly enters the classroom, what changes when students collaborate with algorithms instead of peers? We analyzed 36 undergraduate students learning graph theory through peer collaboration (n=24) or AI assistance (n=12), using discourse analysis to identify interaction patterns shaping learning outcomes. Results reveal a collaboration quality divide: high-quality peer interactions generated curiosity and engagement that AI couldn't match, yet low-quality peer interactions performed worse than AI across dimensions. AI showed a paradoxical pattern, building confidence in knowledge while reducing curiosity and deeper engagement. Interaction quality emerged from dynamic patterns rather than individual traits, with early discourse markers predicting outcomes. Students treated AI as a transactional information source despite its collaborative design, revealing fundamental differences in human versus algorithmic engagement. Our findings suggest AI in education need not replace peer learning but can recognize struggle and support both peer and AI interactions toward productive learning experiences.

## Introduction

Educational technology adoption has accelerated dramatically, with AI-powered tutoring systems now reaching millions of students worldwide. As AI assistants become more sophisticated, they increasingly substitute for peer interaction in educational settings. This rapid integration raises fundamental questions about how AI changes the nature of learning, particularly the social dimensions that have long been central to education.

Vygotsky's foundational work on the Zone of Proximal Development emphasized that learning is inherently social, occurring through interaction with more- and less- knowledgeable others who provide scaffolding within the learner's developmental range (Vygotsky, 1978). Decades of research demonstrate that peer collaboration enhances learning through multiple mechanisms that extend beyond information exchange. When learners explain concepts to peers, they improve their understanding, build secondary skills, and strengthen retention; this phenomenon is extensively documented in learning sciences research (Chi et al., 2001; Roscoe & Chi, 2007). Socially shared regulation, where groups collectively monitor and adjust their learning strategies, predicts both learning process quality and learning outcomes in collaborative settings (Järvelä & Hadwin, 2013; Malmberg et al., 2015). These social processes foster curiosity, motivation, and deeper engagement with material, extending beyond material transfer.

These established social learning mechanisms raise critical questions as AI systems increasingly substitute for peer interaction. This substitution aligns with broader trends in social preferences: recent studies document young people increasingly choosing AI companions over human relationships, citing less judgment, greater control, and reduced social anxiety as key factors (Pentina et al., 2023). In educational contexts specifically, students who regularly use AI tutors report feeling less connected to human teachers and participate less in class discussions (Nazaretsky et al., 2022; Kim et al., 2024). These changes occur against a backdrop of rising loneliness among adolescents and young adults, now identified as an emerging public health crisis (Cacioppo et al., 2018; Twenge et al., 2021), with profound implications for cognitive development and academic achievement.

Consider two scenarios from our research. In one, two students puzzle over an applied graph theory problem together, suggesting half-formed ideas, speaking hesitantly, until one suggests, "Wait, what if we draw it like a map first?" This spark of understanding ripples through their conversation, leading to deeper exploration together. In another scenario, a student works with an AI assistant that probes the student for their own understanding, while also guiding with clear, accurate explanations of centrality measures and weighting. The student gains greater confidence in the terminology but never ventures beyond the assigned problems. Which scenario better serves learning? Our results suggest that the answer is neither simple nor universal.

This study examines these questions through detailed discourse analysis of collaborative learning sessions, comparing peer-peer and human-AI interactions. We analyzed 36 undergraduate students learning graph theory, using computational methods to identify specific discourse patterns that shape learning experiences. We discover a collaboration quality divide that challenges assumptions about both modalities: while high-quality peer interactions generate curiosity and engagement that AI cannot match, low-quality peer collaborations perform worse than AI across multiple dimensions. Even identical AI behavior produces divergent outcomes depending on how students engage with it, suggesting that interaction quality emerges from dynamic patterns rather than the capabilities of either party alone.

Our analysis reveals that specific discourse features, such as giving explanations, question-asking patterns, and turn balance, strongly predict whether a learning session will be perceived positively. These patterns appear early enough to enable real-time intervention, pointing toward a future where one role of AI is not to replace peer collaboration but to detect and support it when needed. By identifying which collaborations need support and understanding the complementary affordances of peers (sparking curiosity, generating interest) and AI (consistent knowledge delivery, procedural confidence), we provide grounding for designing systems that preserve the best of social learning while leveraging AI's unique strengths.

## Related Work

### Understanding Human-Human and Human-AI Collaboration

Researchers in CSCL and beyond learning research have begun exploring a variety of approaches comparing human-human collaboration (HHC) with human-AI collaboration (HAI). Tong et al. (2025) found that while both configurations improved physics problem-solving, HHC showed larger effect sizes than HAI, suggesting that peers offer benefits beyond correct answers. In a meta-analysis, Vaccaro et al. (2024) investigated patterns across different human-AI configurations, finding that human-AI teams underperform on decision tasks but excel at content creation, highlighting the nuances in shaping collaborative effectiveness.

Within learning contexts specifically, research examines the experiences of AI-integrated experiences across dimensions. Järvelä et al. (2025) found differences in perceived agency and ownership between peer and AI co-authors. Hernández-Leo et al. (2025) explored how generative AI can improve some types of collaborative workflows while interfering with others. Kosmyna et al. (2025) explored neural and behavioral impacts of AI-supported writing, emphasizing the tradeoff between immediate improvement and long-term learning implications. These comparisons establish precedent for examining HHC and HAI approaches but focus primarily on outcomes rather than interactional processes.

### Interaction Analysis and Real-Time Mediation in CSCL

Productive collaboration has distinct interactional patterns. Balanced participation is associated with deeper learning and better problem-solving (Barron, 2003). Transactive discourse (building on or challenging others' ideas) promotes more complex understanding than parallel thinking (Berkowitz & Gibbs, 1983; Teasley, 1997). Question-asking drives exploratory talk and collective knowledge construction (Mercer & Littleton, 2007). Constructive controversy, social question-asking, and supported uncertainty promote increased levels of curiosity (Sinha et al., 2017).

The broader field of conversation analytics has developed a wealth of approaches for analyzing collaborative discourse, including using recent LLM advances as an analytic tool. CPSCoach 2.0 (D'Mello et al., 2024) demonstrates real-time classification of collaboration quality using multimodal features, enabling adaptive support during group work. The Jigsaw Interactive Agent (Doherty et al., 2025) uses LLMs to scaffold specific group dynamics like respectful disagreement and evidence-based reasoning. Lee et al. (2025) showed that AI-facilitated small-group collaboration through real-time analytics improved both math learning outcomes and reduced learning anxiety. These systems show the technical feasibility of discourse-aware support but are typically evaluated as interventions rather than instruments for understanding baseline differences across learning ecologies. Our research applies related computational techniques comparatively, aiming to understand how collaborative behaviors like curiosity expressions, responsive questioning, and participatory balance manifest differently when one's partner is an algorithm rather than a peer, and how this impacts both learning and social perception outcomes.

## Methods

### Participants and Learning Task

We recruited thirty-six undergraduate students (ages 18-22; 20 female, 12 male, 4 non-binary) from a large research university. The study was approved by our institutional IRB. Participants were recruited through university mailing lists and a behavioral research platform, receiving $25 for participation. Students were randomly assigned to either peer dyads (12 pairs, n=24) or individual AI-assisted learning (n=12). Participants in peer dyads were not acquainted prior to the session, mimicking the format of drop-in study support rather than established classroom relationships.

All sessions were conducted remotely via Zoom to reflect increasingly common online learning environments and to constrain interaction channels for discourse analysis. The learning activity centered on graph theory concepts applied to network design problems, selected for its balance of conceptual understanding and creative problem-solving. This domain requires both knowledge acquisition (understanding graph properties) and application (designing networks), allowing us to observe different interaction dynamics.

Following a brief instructor-led introduction covering basic concepts (vertices, edges, directionality, weighting), participants engaged in individual study. In the peer condition, pairs received complementary learning materials in a jigsaw design: one student studied network structure while the other studied network dynamics. In the AI condition, students received one of these material sets. This design supported comparable knowledge preparation across conditions while creating interdependence in peer pairs.

Students then engaged in a 30-minute collaborative project session designing transportation or communication networks while optimizing for specific constraints (e.g., minimizing connections while maintaining reachability, designing resilience to failure cases, identifying critical nodes). All participants used Miro, an online whiteboard platform, for the design task, after a brief introductory exercise to build comfort with basic drawing tasks on the platform. AI condition participants interacted with a custom chat assistant through a web interface while completing the same design task. At the end of the project phase, students in both conditions were asked to explain their network design and report how they met the task deliverables.

**Figure 1**
*Study Design Overview*

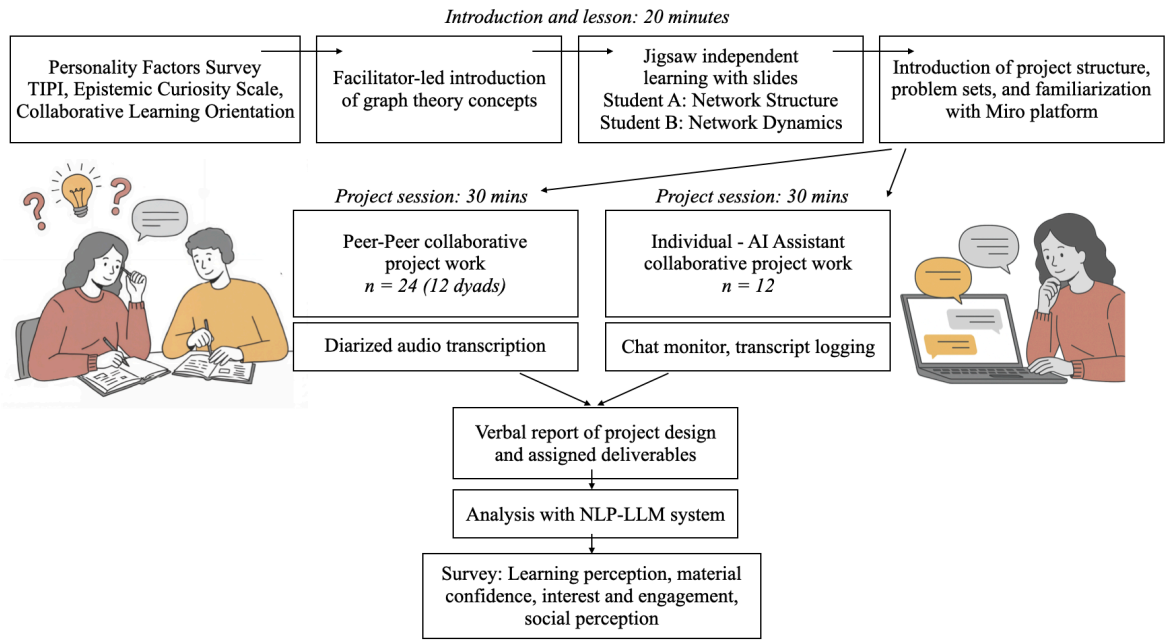

### Chat Assistant Design

The AI assistant was built using the Anthropic API (Claude 3.5 Sonnet) with custom prompting designed to support collaborative learning. The interface presented a conversational chat-bubble interaction accessed through a web browser, with chat data transmitted to a secure server for real-time monitoring. The assistant's prompt design built on established educational chatbot frameworks (Mollick & Mollick, 2023), reflecting a likely real-world use case of existing AI chat interactions. The prompt emphasized qualities including:
- Transparency about being an AI assistant
- Support for learning through problem-solving guidance rather than direct answers
- Encouragement of student reasoning and problem-solving
- Offering structured guidance with clarifying questions

The assistant was provided with knowledge of the lesson materials and individual study content, scoping its support accordingly. Chat transcripts were logged with timestamps for subsequent analysis.

## Data Collection and Analysis

### Pre-session measures
Participants completed baseline assessments including the Ten Item Personality Index (TIPI; Gosling et al., 2003), Epistemic Curiosity Scale (Litman et al., 2003), prior experience with online learning and AI tools, and collaborative learning orientation items to test whether individual differences predicted collaboration outcomes.

### Session recording
All sessions were recorded via screen capture and audio recording. Peer sessions captured both students' verbal exchanges and their shared Miro workspace. AI sessions captured timestamped chat transcripts, Miro workspace activities, and audio/video recording of verbal commentary including the closing presentation.

### Post-session measures
Immediately following the project work, students verbally presented their network designs to the experimenter and completed a 32-item perception survey assessing four dimensions:

- Confidence in learning and material understanding (6 items, *e.g. "I feel that I could teach this material to someone else"*)
- Interest in material (5 items, *e.g. "I might apply these concepts to other projects"*)
- New ideas and curiosity (5 items, *e.g. "I wanted to dive deeper into the material beyond what was taught"*)
- Social perception of learning experience (6 items, *e.g. "Working with my partner/AI helped me learn better than I would have alone"*)

## Discourse analysis
To enable detailed analysis of interaction patterns across conditions, we developed and validated a conversation analytics system that classifies curiosity expressions and interaction features in learning discourse. This methodology allows us to connect micro-level discourse patterns with macro-level learning outcomes.

The system is designed as a hybrid architecture, balancing accuracy, privacy, and computational efficiency. Initial classification uses local NLP classifiers based on semantic patterns for curiosity types (specific, diversive, epistemic, social) and interaction features (questioning, explanation, agreement/disagreement, uncertainty). When local classifiers produce low-confidence results (<0.6 confidence score), the system escalates only decontextualized text segments to an LLM (Anthropic API) for more nuanced classification. In this hybrid approach, approximately 75% of segments are processed locally (preserving privacy and minimizing computational costs), while complex, context-dependent expressions receive more sophisticated analysis.

The classification system was validated against 450 human-annotated conversation segments, with each segment receiving approximately 12 independent human ratings. Inter-rater reliability between the system and human annotators showed moderate to high agreement ($\kappa=0.72$ for curiosity presence, $\kappa=0.76$ for interaction type classification, $\kappa=0.66$ for curiosity subtype classification).

From the classified segments, we computed conversation-level metrics including: *Curiosity expression* rates (normalized instances per conversation), *Social curiosity* ratio (social curiosity / total curiosity expressions), *Explanation* frequency (explanations given per participant), and *Turn-taking* balance.

## Statistical analysis
Our analysis examined three levels of relationships: (1) between-group comparisons of survey measures, (2) within-conversation relationships among discourse features, and (3) correlations between conversation-level discourse metrics and post-session survey responses.

For between-group comparisons of survey responses (5-point Likert scales treated as interval data), we used independent samples t-tests with Cohen's d effect sizes. Groups were formed using median splits within each condition (peer vs. AI) based on collaboration quality perceptions.

To examine relationships between discourse features and learning outcomes, we calculated Pearson correlations at the individual participant level. For peer sessions, discourse features were coded separately for each participant within the dyad (e.g., number of curiosity expressions by participant A, number of explanations given by participant A), with n=24 individual data points. These individual-level discourse metrics were then correlated with each participant's own survey responses. For AI sessions (n=12 individuals), each participant provided one data point.

Given power limitations due to our sample size (n=24 for peer participants, n=12 for AI participants), we report effect sizes alongside significance values with α = .05 for statistical significance. We note that individuals within peer dyads may show correlated outcomes due to their shared interaction experience; however, the exploratory nature of this work and the consistency of patterns across multiple measures suggest meaningful relationships worth further investigation. All analyses were conducted using Python (version 3.10) with scipy.stats and pandas libraries.

## Results

### Collaboration Quality Divide

Analysis revealed substantial variation in collaboration quality both between and within conditions, clustering into distinct quality levels rather than continuous distributions. Participants were grouped by median split on perception/engagement scores within each condition (see Figure 2). For peer-peer interactions, this distinction emerged from surveyed social perception measures (e.g., quality of collaborative experience, partner engagement). For AI interactions, groupings were based on AI engagement perception (e.g., quality of interaction with the assistant). These groupings revealed a pattern we term the "collaboration quality divide."

High-quality peer-peer interactions *(HQ-PP, n=13)* reported the strongest collaborative experiences (M=4.85, SD=0.38), significantly outperforming both low-quality peer-peer collaborations *(LQ-PP, n=11)* (M=3.82, SD=1.25; t(22)=2.58, p=.017, d=1.16) and all AI conditions (M=3.10, SD=1.20; t(23)=4.53, p<.001, d=2.10). This gap between high-quality peer and AI interactions suggests fundamentally different learning experiences rather than incremental differences.

Low-quality peer-peer interactions performed worse than AI participants on multiple measures, challenging assumptions about the inherent value of peer learning. This pattern held across learning confidence (LQ-PP: M=3.18, SD=0.75 vs. AI: M=4.30, SD=0.67; t(21)=-3.62, p=.002, d=-1.56) and interest in the material (LQ-PP: M=3.36, SD=0.92 vs. AI: M=3.75, SD=0.75; t(21)=-1.08, p=.29, d=-0.46).

**Figure 2**
*Participant ratings of peer-learning perception or AI-learning perception*

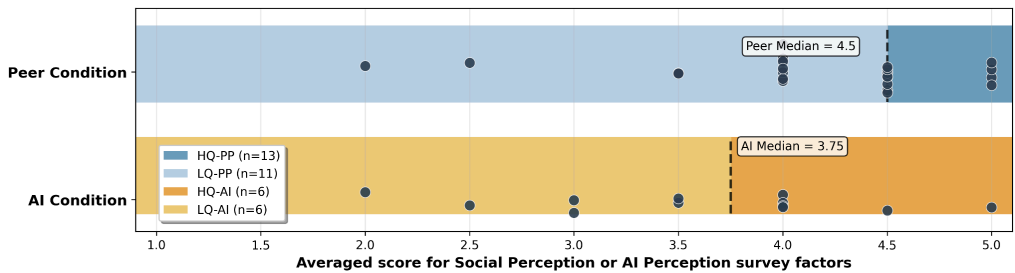

Participants are separated at each condition's median into high quality (HQ) and low quality (LQ) interaction groups per condition.

### Peer-Peer / Individual-AI Differential Effects

Groups showed distinct patterns across different aspects of the learning experience (see Figure 3). For confidence in explaining concepts ("I feel capable of helping others understand the topic"), a surprising pattern emerged: AI participants (M=4.30, SD=0.67) and LQ-PP participants (M=3.18, SD=0.75) reported the highest and lowest confidence respectively (t(21)=-3.62, p=.002, d=-1.56). This suggests that an AI-based interaction can build subject-material confidence even when deeper engagement is limited, perhaps because of the consistency and factual nature of the AI support.

Conversely, for generating new ideas and curiosity ("I found myself asking new questions beyond the lesson material", "I wanted to explore new ideas", "The project sparked deeper interest"), HQ-PP participants (M=4.00, SD=0.91) significantly outperformed AI participants (M=3.10, SD=0.99; t(23)=2.16, p=.043, d=1.00). The LQ-PP exchanges fell between these extremes (M=3.36, SD=0.92), indicating that even struggling peer interactions generate more curiosity than AI exchanges. This "confidence without curiosity" pattern in AI interactions reflects their more transactional nature: students get answers efficiently but miss the generative friction of peer collaboration.

**Figure 3**
*Between-groups comparison across survey factors*

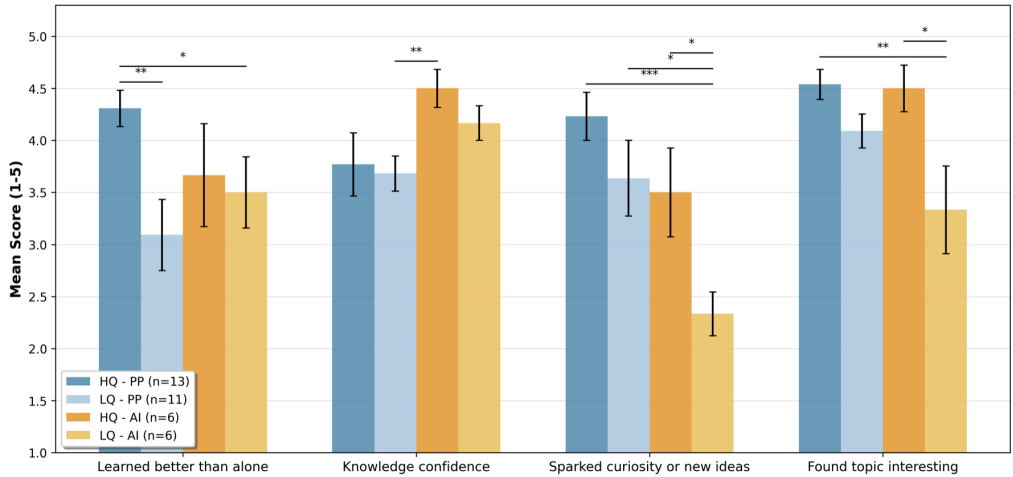

Comparison of four groups across survey measures from categories (L-R): peer/AI learning value, knowledge confidence, curiosity, and interest. From L-R in each category: HQ-PP group, LQ-PP group, HQ-AI group, LQ-AI group. *Note*: *p<.05, **p<.01, ***p<.001

## Discourse Patterns as Predictors

Discourse analysis on peer pairs reveals significant relationships between specific interaction patterns and learning perceptions (Table 1). While our sample size limits statistical power (n=24 for peer condition analyses), the observed correlations suggest meaningful relationships warranting further investigation. These correlations also show consistency with collaborative learning theory: social curiosity and explanation are established mechanisms through which peer learning benefits emerge.

**Table 1**
*Emergent interaction features and correlations to curiosity, interest, and survey factors.*

| Interaction feature | Curiosity / Perception component | r | p |
| --- | --- | --- | --- |
| Question asking | Overall curiosity rate | 0.565 | 0.0041** |
| Question asking | Social curiosity rate | 0.642 | 0.0007*** |
| Explanation giving | Social curiosity rate | 0.439 | 0.035* |
| Social curiosity rate | Asked new questions (Q26) | 0.468 | 0.022* |
| Social curiosity rate | Understanding of concepts (Q21) | 0.439 | 0.035* |
| Explanation giving | Value of collaboration (Q17) | 0.417 | 0.039* |
| Overall curiosity rate | Understanding of concepts (Q21) | 0.381 | 0.062 |

*Note: *p<.05, **p<.01, ***p<.001.* Only the most relevant correlations are shown.

Social curiosity expressions—interest in the partner's perspective or reasoning—showed particularly strong correlations with learning outcomes. Students who expressed more social curiosity reported asking new questions beyond the lesson material (r=0.468, p=.022) and better understanding of graph theory concepts (r=0.439, p=.035). Additionally, explanation-giving behavior positively correlated with perceived value of collaboration (r=0.417, p=.039). The strong correlation between question-asking and social curiosity (r=0.642) likely reflects overlapping constructs in our coding, as social curiosity was frequently enacted through questions about partners' perspectives (e.g., "Why did you do it that way?"), an expected behavior in peer dialogue.

Qualitative analysis of discourse revealed distinct patterns distinguishing high- from low-quality collaborations. Sessions receiving high social perception ratings displayed specific discourse markers early in interactions: frequent questions about partner reasoning ("Why did you connect those nodes that way?"), explicit building on partner contributions ("Like you mentioned about redundancy, but what if we also considered..."), and verbal acknowledgment of learning ("Oh, I hadn't thought about that").

Figure 4 illustrates these contrasting patterns. High-quality conversations showed balanced turn-taking, diverse interaction types (questioning, explaining, disagreement), and greater presence of social curiosity expressions. Low-quality conversations, despite sometimes having balanced turn lengths, consisted primarily of agreement or neutral statements without deeper engagement; participants were often agreeable but not genuinely collaborative.

**Figure 4**

*Comparison between HQ-PP and LQ-PP conversations*

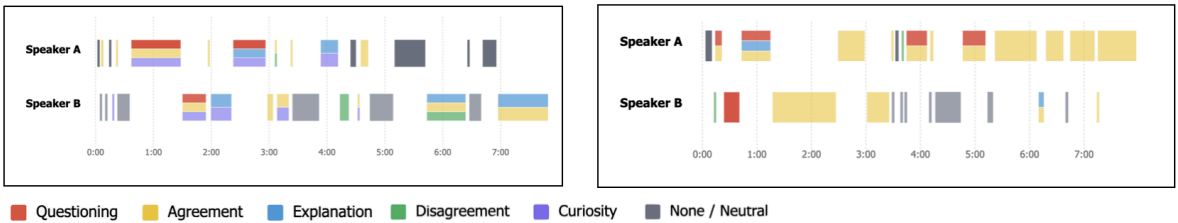

Discourse patterns in representative peer conversations. Horizontal bars show speaker turns over time (width = turn length) with two rows per conversation representing each speaker. Colors indicate coded interaction features (see legend). Vertically stacked colors within a single bar indicate multiple features in one turn; for example, "Yeah, I think you're right. Look, we can also do that with an undirected graph like this, though. How should the shuttle work here?" would be identified as containing *agreement*, *explaining,* and *questioning*.)

## Turn-Taking and Interaction Asymmetry

We also noted striking differences in turn taking, both between positive and negative perception groups per condition, as well as between modalities. HQ-PP dyads demonstrated more balanced contributions (turn ratio ≈ 0.45-0.55), while LQ-PP students showed greater imbalance (turn ratios ranging 0.20-0.80). AI sessions showed extreme asymmetry: students provided minimal responses (M=5 words per turn) compared to AI's lengthier explanations (M=27 words).

Qualitatively, the nature of these interactions differed as well. Students in AI conditions frequently ignored or resisted collaborative prompts from the assistant, treating it as an information source rather than a collaborative partner. One participant explicitly stated: "stop asking me questions, I just want to know what you think." Throughout AI sessions, we observed students providing brief confirmations ("okay," "got it") rather than engaging with the assistant's attempts at dialogue.

These patterns were also clear in the final presentations. Students in AI conditions frequently ignored or resisted collaborative prompts from the assistant. Analysis of the project presentation transcripts showed that a majority of AI participants (9 out of 12) used exact terminology from assistant responses in their presentations, also using more technical language ("This uses adaptive routing for high-capacity paths"), mirroring the AI assistant. Several students, when asked for details about their decision making, acknowledged their more passive role: "The chat suggested that I include that, and it seemed like a good idea".

In contrast, peer participants demonstrated more ownership of their solutions. Their presentations emphasized the collaborative process: "So, first we thought about having a central hub, but then we realized that we should use a triangulated system..." Even during presentations, they negotiated who would explain what: "Do you want to describe how we did the weighting?" Peer participants used less technical jargon, instead explaining concepts in their own words. This distinction between passive reception and active construction reflects our discovery: AI participants reported higher confidence in explaining concepts despite showing lower curiosity and engagement.

## Interaction Quality is Emergent, Not Individual

Critically, interaction quality emerged from dynamic patterns rather than individual traits. Personality measures collected pre-session (TIPI dimensions, curiosity orientation, collaboration preference, topic interest) showed no significant correlations with any outcome variables (all p>.20). This held true across both conditions, suggesting that collaboration success depends on what unfolds between partners, not who they are individually.

The AI condition provided additional evidence for this trend. Despite receiving nearly identical chatbot behavior, participants diverged into the HQ-AI group (n=6) and LQ-AI group (n=6) based on their interaction patterns. The high-engagement group maintained higher question frequency and attempted to build on AI suggestions, while the low-engagement participants adopted mostly transactional patterns, ignoring questions and prompts from the AI assistant. These behavioral differences predicted perception outcomes as strongly as the peer-condition divide.

## Discussion

### Complementary Affordances and the Nature of Collaboration

Our findings reveal that the value of peer learning depends critically on interaction quality, not inherent superiority of a single mode. The collaboration quality divide demonstrates peer learning as high-risk, high-reward: when effective, it significantly outperforms AI on engagement and curiosity metrics, but when it fails, students might be better served by AI assistance. This variability emerges from dynamic relational patterns rather than individual traits, as evidenced by the lack of personality correlations.

The effects suggest fundamentally different affordances. AI excels at consistent knowledge delivery and building procedural confidence; students felt capable of explaining concepts despite showing less engagement. This "confidence without curiosity" pattern reflects transactional human-AI interaction where students receive answers efficiently but miss the generative friction of navigating uncertainty together. Notably, students naturally adopted these different interaction modes. They didn't treat AI as a collaborator despite its attempts at Socratic dialogue; they used it as a tool, suggesting systems should embrace rather than mask their tool nature.

### Theoretical Implications

These initial findings point toward extending Vygotsky's ZPD concept into AI-mediated contexts. While AI provides scaffolding, it may not activate the same socio-cognitive mechanisms as peer interaction. High-quality peer discourse showed a collective navigation of uncertainty that sparked curiosity and deeper engagement. The strong correlation between social curiosity and learning outcomes suggests that curiosity about others' thinking processes may be as important as content curiosity itself, a dimension that is often absent in AI interactions.

Our finding that interaction quality emerges from dynamic patterns rather than individual traits also points to the collaborative dynamics of aptitude. Rather than students being inherently "good" or "bad" at collaboration, success depends on the interaction dynamics between specific partners in specific contexts, as seen with identical AI behavior producing divergent outcomes based on student engagement patterns.

### Design Implications and Limitations

The discourse markers that suggest interaction quality – asking questions, showing social interest, giving explanations – appear early enough for real-time intervention. Rather than AI replacing peers, it could monitor collaborations and provide support when struggling patterns emerge (extreme turn imbalance, low questioning rates). Future systems might combine modalities: leveraging peer interaction for curiosity generation and AI for knowledge consolidation when peer dynamics falter. Translating these findings into practice will require close partnership with educators to develop systems that are both effective for learning and feasible to implement in real classroom contexts.

We acknowledge several limitations within our research. Sessions were single encounters lasting approximately one hour; longer-term collaborations might show different patterns, particularly as relationships develop and students learn to work together. The specific task (graph theory network design) and remote format may not generalize to other learning contexts or face-to-face settings. Our university student sample may not represent younger learners or those with different AI expectations. Additionally, the discourse analysis system's moderate inter-rater reliability ($\kappa = 0.66$-$0.76$) reflects both technical limitations and inherent subjectivity in identifying curiosity expressions. Our sample sizes for AI subgroups were small (n=6 for both positive and negative perception conditions), though they showed initial consistent patterns across multiple analyses.

### Conclusion

As educators integrate AI into classrooms, the question isn't whether it should replace peer learning but how it can strengthen it. By identifying early markers of collaboration quality and understanding the complementary affordances of peers (sparking ideas, generating interest and curiosity) and AI (consistent knowledge delivery), we can design environments that preserve social learning benefits while leveraging AI's strengths. Our findings suggest that effective inclusion of AI in education goes beyond seeing human or AI collaborators as uniformly beneficial or detrimental, but rather detecting and nurturing quality interactions wherever they emerge.

## Acknowledgments

This research was supported by a fellowship from the MIT Morningside Academy for Design. We thank the MIT Behavioral Research Lab for their support with research and participant management.